\documentclass{PoS}

\usepackage{lscape}
\usepackage{afterpage}
\usepackage{amsmath}
\usepackage[square,comma,numbers,sort&compress]{natbib}

\newcommand*{\no}{\noindent}
\newcommand*{\bea}{\begin{eqnarray}}
\newcommand*{\eea}{\end{eqnarray}}
\newcommand*{\be}{\begin{equation}}
\newcommand*{\ee}{\end{equation}}

\newcommand*{\pref}[1]{(\ref{#1})}

\newcommand{\bma}{\begin{pmatrix}}
\newcommand{\ema}{\end{pmatrix}}

\title{On the phase diagram and the singlet scalar channel in Yang-Mills-Higgs theory}

\ShortTitle{On the singlet scalar channel in Yang-Mills-Higgs theory}

\author{\speaker{Axel Maas}\thanks{Supported by the DFG under grant numbers MA 3935/5-1, MA/3935/8-1 (Heisenberg program), and GK 1523/2.}\\
        University of Graz, Institute for Physics, Universit\"atsplatz 5, A-8010 Graz, Austria\\
        E-mail: \email{axel.maas@uni-graz.at}}
        
\author{Tajdar Mufti\thanks{Supported by the DFG graduate school GK 1523/1 and GK 1523/2 and under grant number MA 3935/5-1}\\
	Institute of Theoretical Physics, Friedrich-Schiller-University Jena, Max-Wien-Platz 1, D-07743 Jena, Germany\\
        E-mail: \email{tajdar.mufti@uni-jena.de}}

\abstract{Yang-Mills-Higgs theory is quite a remarkable theory in that it shows very different behaviors without phase transitions. It is dominated by the Brout-Englert-Higgs mechanism in some domain of the phase diagram, while it is essentially QCD-like in another. It is expected that albeit there is no qualitative difference, there are substantially quantitative differences throughout the spectrum. This is investigated using lattice theory for the case of the scalar singlet channel for more than a hundred different points in the phase diagram. It is found that the results deviate partly substantially from the expectations in some cases, but in others justify the picture of a weakly interacting theory - even in cases of rather strong interactions at the ultraviolet cutoff.}

\FullConference{The 32nd International Symposium on Lattice Field Theory,\\
		23-28 June, 2014\\
		Columbia University New York, NY}

\begin{document}

\section{Introduction}

Yang-Mills-Higgs theory, i.\ e.\ an SU(2) Yang-Mills theory coupled to a complex doublet of scalar particles, is both conceptually and phenomenologically relevant. The phenomenological relevance comes about since it is just the Higgs sector, though with degenerate $W$ and $Z$ bosons because of the absence of QED, of the standard model. Though without the fermions it is substantially weaker interacting \cite{Maas:2013aia}, many qualitative statements should hold true\footnote{We assume here than the possible triviality of the theory is a sufficiently small effect below the cutoff, here simulated by the lattice cutoff $a^{-1}$ introduced by the lattice spacing $a$, to not affect our results.}.

In particular, it should be possible to cross-check, whether the expectation about this theory, especially as motivated by perturbation theory \cite{Bohm:2001yx}, should hold true. These are that the theory shows two distinct behaviors, though they are necessarily not separated by a phase transition \cite{Osterwalder:1977pc,Fradkin:1978dv}. In the one domain, at strong gauge coupling, the theory should exhibit a QCD-like behavior. This has been confirmed, e.\ g., by the observation of a non-vanishing intermediate-distance string tension \cite{Knechtli:1998gf}, just like in QCD. As a consequence, the spectrum should show Regge trajectories, and other features known from QCD, which are not (directly) related to chiral symmetry.

On the other hand, there should be a region at weak gauge coupling where a Brout-Englert-Higgs (BEH) effect operates. Naively, this region may be either at weak or at strong Higgs-self-coupling. In the latter case, the Higgs becomes heavy, much heavier than the gauge bosons. In the perturbative domain, the mass of the Higgs should be, essentially, free to choose. Here, to leading order, it does not matter whether continuum or lattice perturbation theory is used. These results depend essentially on the classical action, i.\ e.\ the action at tree-level, and hence at the cutoff. But the usual argument is that in the weak interaction domain this should remain true, even if there are quantum corrections to the parameters.

Of course, this is not the full truth, as also in the BEH domain the observables must be gauge-invariant states \cite{Frohlich:1980gj}, i.\ e.\ bound states from the point of view of field theory. That perturbation theory still adequately describes experiments is due to the Fr\"ohlich-Morchio-Strocchi (FMS) effect \cite{Frohlich:1980gj,Frohlich:1981yi}, confirmed in lattice calculations \cite{Maas:2013aia,Maas:2012tj}. The FMS mechanism shows that, in suitable gauges, a description in terms of either the physical degrees of freedom or the elementary degrees of freedom will yield the same results, up to corrections of order of the Higgs fluctuations, being very small in the standard model. Essentially, there are bound states dual to the elementary particles, having the same mass to leading order. This justifies perturbation theory. But then, the logical possibility exists that there may be internal excitations, which could in principle be observable, as well as bound states in other quantum number channels \cite{Maas:2012tj}.

Indications that the so sketched picture may not be entirely correct have already been found long ago \cite{Evertz:1985fc,Langguth:1985dr}: It was not possible to push the $W$ boson mass - or, more appropriately said, the one of its dual gauge-invariant bound state - above the mass in the corresponding scalar channel while maintaining the BEH effect. To understand this behavior, and also to investigate whether other parts of the naive picture may not be entirely correct, we study here the physical spectrum in a large part of the phase diagram. In total, we have investigated about 115 different points, using the methods described in \cite{Maas:2013aia,Maas:unpublished3}.

\section{Phase diagram}

The first interesting question is, whether the observation in \cite{Evertz:1985fc,Langguth:1985dr} is correct, and under which conditions the physics is QCD-like or BEH-like. Of course, since there is no phase transition between the two regions, this not a well-posed question. But since the presence or absence of the BEH-effect is anyhow dependent on the chosen gauge \cite{Caudy:2007sf}, the question in itself is dependent on the choice of gauge. We will therefore use the volume-dependence of the relative alignment of the Higgs field $\langle(\sum\phi)^2\rangle$ \cite{Caudy:2007sf} in non-aligned minimal Landau gauge \cite{Maas:2012ct} as an indication of the BEH effect. In addition, we monitor the effective mass of the $W$ boson at long times. The BEH effect requires this to be an (almost) ordinary mass, while in the QCD-like domain, it is not well-defined \cite{Maas:2011se,Maas:2013aia}. One could argue that the intermediate-distance string tension would be also a helpful quantity. However, if the distance over which an almost linear rise appears shrinks and shrinks, it will at some point not longer be distinguishable from the remaining effects, even if there still is a remnant.

To compare these quantities to the observable spectrum, we calculated the level spectrum in the dual channels to the Higgs, being the $0^+$ custodial singlet, and the $W$, being the $1^-$ custodial triplet \cite{Frohlich:1981yi}, essential with the methods described in \cite{Maas:2013aia}. However, in addition, we used an enlarged operator basis and preconditioning, as will be described elsewhere \cite{Maas:unpublished3}. This basis included up to 10 operators in the $0^+$ channel and 8 in the $1^-$ channel, depending on the statistical fluctuations. In both cases two-particle scattering states, in the $1^-$ channel even three-particle states, where part of the basis. In this section, only the ground-state masses are relevant, and have been obtained after the variational analysis with two-$\cosh$ fits. All the masses were obtained on 24$^4$ lattices. Again, details will be presented elsewhere \cite{Maas:unpublished3}.

\begin{figure}
\centering
\includegraphics[width=0.5\linewidth]{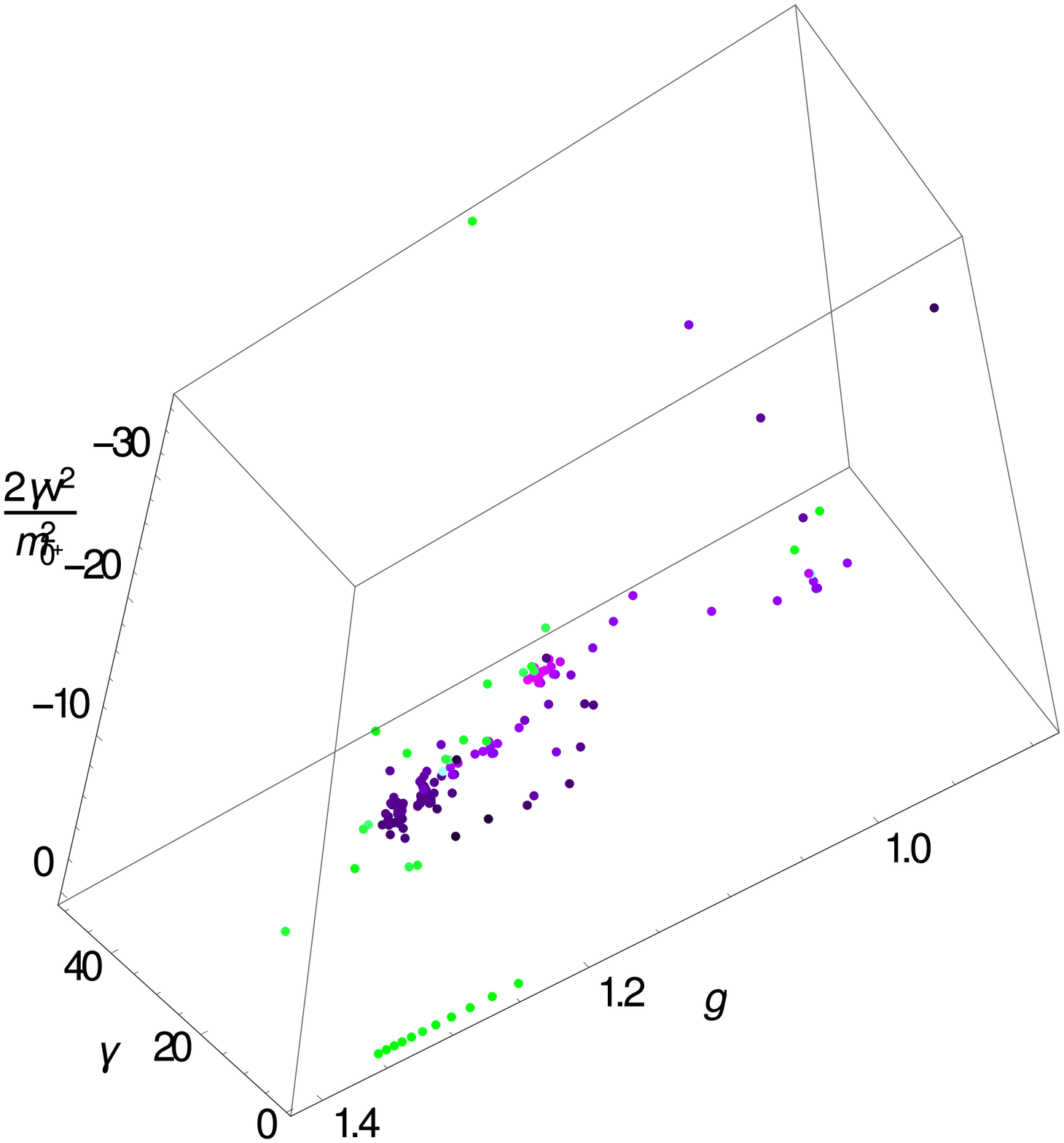}\includegraphics[width=0.5\linewidth]{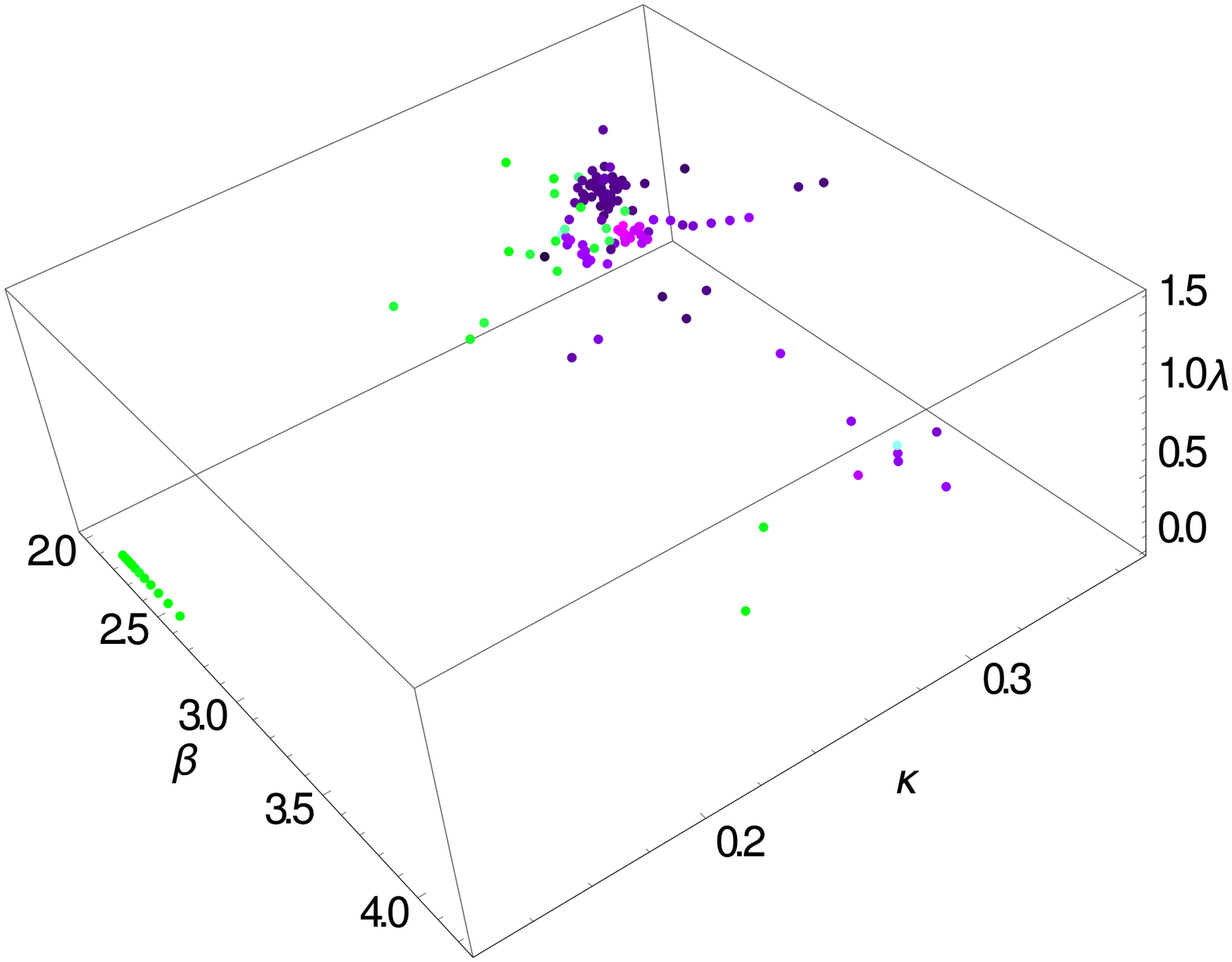}
\caption{\label{lcp}The left panel shows the phase diagram of the Yang-Mills-Higgs theory as a function of bare gauge coupling, Higgs 4-point coupling, and the bare Higgs mass in units of the $0^+$ singlet mass. Green points are QCD-like, and purple points are BEH-like. The lighter the points, the smaller is the lattice spacing. The right-hand plot shows the same in terms of the lattice bare parameters of inverse gauge coupling, hopping parameter, and four-Higgs coupling, see \cite{Maas:2013aia} for their relation to the continuum parameters.}
\end{figure}

The results are shown in figure \ref{lcp}. The points in the phase diagram have been classified using the various quantities listed above. Using the lattice couplings, the BEH phase persists at all gauge couplings, for sufficiently large hopping parameter and not too large Higgs self-coupling. Translated into the continuum, the BEH phase is driven like a wedge into the QCD-like region, up to rather large gauge couplings, significantly above 1. However, in contrast to the perturbative situation \cite{Bohm:2001yx}, the QCD-like domain persists even at substantially bare negative mass-squared for the Higgs. In addition, the Higgs self-coupling must also be small enough for the BEH effect to become operational, and otherwise the theory is driven into the QCD-like domain.

\section{Levels in the singlet scalar channel}

\afterpage{\begin{landscape}
\begin{figure}[p]
\includegraphics[width=\linewidth]{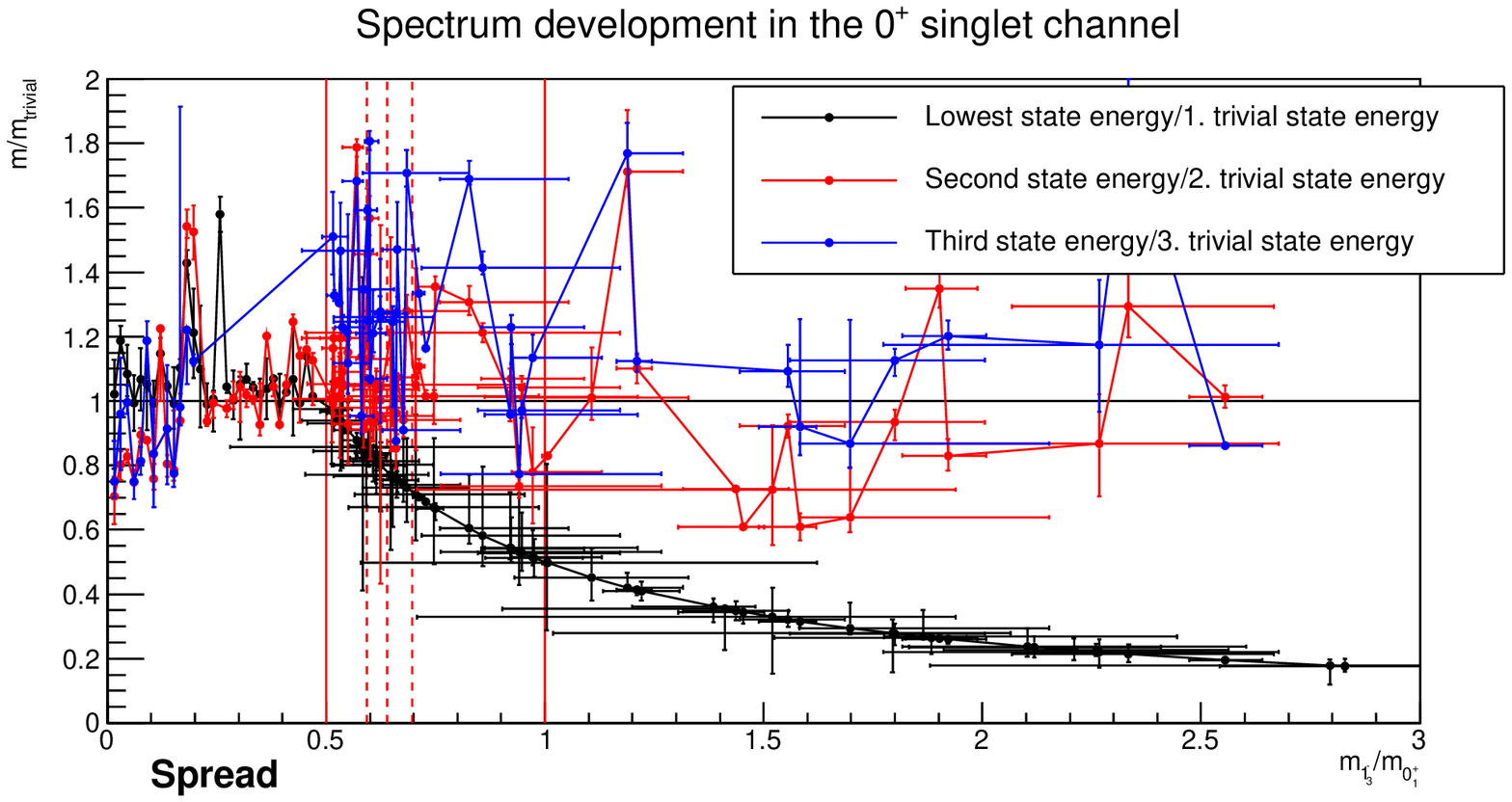}
\caption{\label{m0}The three lowest energy levels in the $0^+$ singlet channel, as a function of the ratio of the ground state mass in this channel and the ground state mass in the $1^-$ triplet channel. The word 'Spread' indicates that the ratio does not make any sense anymore once the $0^+$ singlet becomes unstable to the decay into the $1^-$ triplet channel, and above this threshold the points would all coalesce at the value 0.5, but are spread according to their lattice spacing, the smallest lattice spacing to the lowest values. The $y$ axis is the mass of the $0^+$ state divided by the trivial mass, as discussed in the text. The dashed vertical line delineate a 125$\pm$10 GeV band, corresponding to the physical Higgs mass. Not in all cases three levels could be reliably identified. Furthermore, points with an average energy greater than $\pi/2$ have been omitted.}
\end{figure}
\end{landscape}}

To better understand the results from the phase diagram, it is useful to investigate the lowest levels in the $0^+$ channel. These are shown in figure \ref{m0}. Plotted are the three lowest energy levels, wherever they were possible to identify, and still somewhat reasonably small compared to the lattice spacing. They are plotted against the ratio of the ground-state masses in the $0^+$ singlet to the $1^-$ triplet channel. At sufficiently large mass of this ground-state, however, the $0^+$ singlet can decay into two triplets. Fixing the mass of the $1^-$ triplet to 80 GeV, this would be at 160 GeV. It then makes no longer sense to normalize the results such, as the lowest level becomes a scattering state, and all values would coincide at 0.5. To avoid this cluttering up, the points are spread evenly according to their lattice spacing, determined from the fixed $1^-$ triplet channel, the finest lattice closest to zero.

Shown as a function are the three lowest energy levels in the $0^+$ singlet channel. They are normalized to the expected lowest three trivial states. Since it appears that always either the $0^+$ singlet or $1^-$ triplet channel are the lowest channel \cite{Maas:unpublished3}, these trivial states are determined in the following way:

\begin{itemize}
\item The lowest state is normalized to twice the mass of the $1^-$ triplet ground state. Hence, it should become one, if the mass of the $0^+$ crosses the decay threshold. Below the decay threshold, this will depend on the mass of the $1^-$ triplet, but must be a strictly decreasing function given the $x$-axis.
\item The normalization of the next two states depends upon whether the decay threshold is crossed for the ground state. The second state is normalized to twice the lighter of either the $0^+$ singlet ground-state mass or the $1^-$ triplet ground-state mass, as long as the ground-state in the $0^+$ singlet channel did not yet cross the decay-threshold, as then it could be the first trivial level. If it has crossed, then it is normalized to the first state with the minimum kinetic energy in an $s$-wave, i.\ e.\
\be
m_\text{trivial}=2\sqrt{m_{1^-_3}^2+\left(\frac{2\pi}{N}\right)^2}\label{mt1}
\ee
\item Correspondingly, the second state is normalized to the value \pref{mt1} below the decay threshold, and to the trivial state with two units of momenta above the decay threshold.
\end{itemize}
\no This normalization ensures that if all states should be trivial, then all curves would always be one, within statistical and systematical errors.

The results show a different picture, however. There is indeed a substantial range in which there is a non-trivial ground state, as expected from the previous section. This state is stable inside this theory, and conforms to the picture of a Higgs. After crossing the decay threshold, it indeed becomes trivial, being just a scattering state.

The situation is somewhat different for the higher levels, though their uncertainties are substantially larger. The first surprise is that in the QCD-like region all results cluster essentially around one, in both cases. This implies that there appears to be no additional states below the decay threshold of two $0^+$ singlet ground state particles in an $s$-wave in the $0^+$ singlet channel. That is quite surprising, since in a strongly interacting theory, especially one which should be able to sustain Regge trajectories, the naive expectation is that there should be a number of states with level spacings similar as in QCD. Hence, there should be several states far below one. This is not observed, though the uncertainties, as stated are large - shown are only one-$\sigma$ statistical errors. This will require a substantially better investigation.

In the range where the $0^+$ singlet has one to two times the $1^-$ triplet mass, i.\ e.\ in the physical range of a light Higgs, the second state is compatible with a trivial state. The third state comes up somewhat higher, but this would then rather require to find the second trivial state rather than to overinterpret it as non-trivial state. The result is nonetheless interesting. Such a situation is expected at weak coupling, and indeed seen in lattice calculations at substantially weaker bare couplings than the ones investigated here \cite{Wurtz:2013ova}. Here, however, both the four-Higgs coupling and the gauge coupling are comparatively strong at the ultraviolet cutoff, as can be seen in figure \ref{lcp}, and deviations could have been expected. Indeed, some are seen, but in other channels, as will be presented elsewhere \cite{Maas:unpublished3}. Unfortunately, this also implies that if there is something like an internally excited Higgs, as could be speculated given the FMS mechanism \cite{Maas:2012tj}, it will be very hard to find.

The last interesting point is beyond the decay threshold. At (relatively) large lattice spacings (i.\ e.\ close to, but smaller than, 1/2 in figure \ref{m0}), the second level is, for all intents and purposes, trivial, while the third state cannot be reliably extracted, as it is too heavy. From perturbation theory it is expected that a reasonably stable Higgs above the decay threshold should exist for quite some mass range \cite{Bohm:2001yx}. Here, no evidence is seen for coarse lattices. On finer lattices, the second and third state seem to come down. However, since also the volume becomes smaller, this may just be a finite volume effect, and more refined calculations, e.\ g.\ using L\"uscher's method \cite{Luscher:1991cf} on multiple volumes, will be required to clarify this. If also there all states are trivial, this could lead again to a possible contradiction to the naive assumption about the structure of the theory.

\section{Conclusions}

After establishing with full non-perturbative calculations support for the FMS mechanism, and hence a rather clean field-theoretical interpretation of what is measured at the LHC and its predecessors, understanding the implications for the Higgs sector is the next interesting challenge. As was shown throughout, not all results confirm with the naive, mostly perturbative, expectations about how this theory behaves for the various values of the couplings at the cutoff. Of course, triviality and lattice artifacts may cloud the view, and hence further careful investigations will be required to find a full picture. But if these results are confirmed than even rather weakly-interacting theories with an operative BEH effect hold non-perturbative surprises.

Even if is confirmed that the Higgs sector of the standard model holds no experimentally relevant non-perturbative deviations from the perturbative calculations, this is not the end of the story. Similar sectors appear throughout many beyond-the-standard-model scenarios. Given the results, and the intricacies of the FMS mechanism, in such cases sizable deviations from the perturbative picture can arise. An example is the 2-Higgs-doublet model, see \cite{Maas:2014nya}, where a difference may arise due to the mismatch of the custodial group and the gauge group. It is therefore urgently necessary not only to finalize the assessment of the non-perturbative corrections to the standard model, but also to start understanding other theories with less simpler Higgs sectors than in the standard model.

\bibliographystyle{bibstyle}
\bibliography{bib}

\begin{thebibliography}{10}

\bibitem{Maas:2013aia}
A.~Maas and T.~Mufti,
\newblock JHEP {\bf 1404}, 006 (2014), 1312.4873.

\bibitem{Bohm:2001yx}
M.~B\"ohm, A.~Denner, and H.~Joos,
\newblock {\em {Gauge theories of the strong and electroweak interaction}}
  (Teubner, Stuttgart, 2001).

\bibitem{Osterwalder:1977pc}
K.~Osterwalder and E.~Seiler,
\newblock Annals Phys. {\bf 110}, 440 (1978).

\bibitem{Fradkin:1978dv}
E.~H. Fradkin and S.~H. Shenker,
\newblock Phys. Rev. {\bf D19}, 3682 (1979).

\bibitem{Knechtli:1998gf}
ALPHA collaboration, F.~Knechtli and R.~Sommer,
\newblock Phys. Lett. {\bf B440}, 345 (1998), hep-lat/9807022.

\bibitem{Frohlich:1980gj}
J.~Fr\"ohlich, G.~Morchio, and F.~Strocchi,
\newblock Phys.Lett. {\bf B97}, 249 (1980).

\bibitem{Frohlich:1981yi}
J.~Fr\"ohlich, G.~Morchio, and F.~Strocchi,
\newblock Nucl.Phys. {\bf B190}, 553 (1981).

\bibitem{Maas:2012tj}
A.~Maas,
\newblock Mod.Phys.Lett. {\bf A28}, 1350103 (2013), 1205.6625.

\bibitem{Evertz:1985fc}
H.~G. Evertz, J.~Jersak, C.~B. Lang, and T.~Neuhaus,
\newblock Phys. Lett. {\bf B171}, 271 (1986).

\bibitem{Langguth:1985dr}
W.~Langguth, I.~Montvay, and P.~Weisz,
\newblock Nucl. Phys. {\bf B277}, 11 (1986).

\bibitem{Maas:unpublished3}
A.~Maas and T.~Mufti,
\newblock unpublished.

\bibitem{Caudy:2007sf}
W.~Caudy and J.~Greensite,
\newblock Phys. Rev. {\bf D78}, 025018 (2008), 0712.0999.

\bibitem{Maas:2012ct}
A.~Maas,
\newblock Mod. Phys. Lett. {\bf A27}, 1250222 (2012).

\bibitem{Maas:2011se}
A.~Maas,
\newblock Phys.\ Rep.\ {\bf 524}, 203 (2013), 1106.3942.

\bibitem{Wurtz:2013ova}
M.~Wurtz and R.~Lewis,
\newblock Phys.Rev. {\bf D88}, 054510 (2013), 1307.1492.

\bibitem{Luscher:1991cf}
M.~Luscher,
\newblock Nucl.Phys. {\bf B364}, 237 (1991).

\bibitem{Maas:2014nya}
A.~Maas,
\newblock (2014), 1410.2740.

\end{thebibliography}

\end{document}